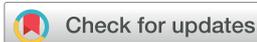

# Sensitivity of Kβ mainline X-ray emission to structural dynamics in iron photosensitizer†

Johanna Rogvall,[a] Roshan Singh,[b] Morgane Vacher [ID] *[c] and Marcus Lundberg [ID] *[d]

Photochemistry and photophysics processes involve structures far from equilibrium. In these reactions, there is often strong coupling between nuclear and electronic degrees of freedom. For first-row transition metals, Kβ X-ray emission spectroscopy (XES) is a sensitive probe of electronic structure due to the direct overlap between the valence orbitals and the 3p hole in the final state. Here the sensitivity of Kβ mainline (Kβ$_{1,3}$) XES to structural dynamics is analyzed by simulating spectral changes along the excited state dynamics of an iron photosensitizer [Fe$^{II}$(bmip)$_2$]$^{2+}$ [bmip = 2,6-bis(3-methyl-imidazole-1-ylidine)-pyridine], using both restricted active space (RAS) multiconfigurational wavefunction theory and a one-electron orbital-energy approach in density-functional theory (1-DFT). Both methods predict a spectral blue-shift with increasing metal–ligand distance, which changes the emission intensity for any given detection energy. These results support the suggestion that the [Fe$^{II}$(bmip)$_2$]$^{2+}$ femtosecond Kβ XES signal shows oscillations due to coherent wavepacket dynamics. Based on the RAS results, the sensitivity to structural dynamics is twice as high for Kβ compared to Kα, with the drawback of a lower signal-to-noise ratio. Kβ sensitivity is favored by a larger spectral blue-shift with increasing metal–ligand distance and larger changes in spectral shape. Comparing the two simulations methods, 1-DFT predicts smaller energy shifts and lower sensitivity, likely due to missing final-state effects. The simulations can be used to design and interpret XES probes of non-equilibrium structures to gain mechanistic insights in photocatalysis.

## 1 Introduction

Formation of an electronically excited state after photon absorption leads to non-equilibrium dynamics along one or several relaxation pathways. The outcome of these photochemical and photophysical processes depend on the coupling between nuclear and electronic degrees of freedom. X-ray spectroscopy is a powerful tool for studies of non-equilibrium processes, with a time resolution down to femtoseconds, through the combination of laser pumps and ultrafast X-ray sources.[1,2] The X-ray probe is element specific and does not interfere with the optical pump due to the large difference in wavelength.

[a] *Department of Chemistry-Ångström Laboratory, Uppsala University, SE-751 20 Uppsala, Sweden*
[b] *Physical and Theoretical Chemistry Laboratory, University of Oxford, South Parks Road, Oxford, OX1 3QZ, UK*
[c] *Nantes Université, CNRS, CEISAM, UMR 6230, F-44000 Nantes, France. E-mail: morgane.vacher@univ-nantes.fr*
[d] *Department of Chemistry-Ångström Laboratory, Uppsala University, SE-751 20 Uppsala, Sweden. E-mail: marcus.lundberg@kemi.uu.se*
† Electronic supplementary information (ESI) available: Additional figures and tables (PDF). *XYZ* coordinates of all structures (ZIP). See DOI: https://doi.org/10.1039/d2cp05671b

An important class of photocatalytic systems are those containing transition metals. For ultrafast studies of coordination complexes, both Kα (2p → 1s) and Kβ (3p → 1s) X-ray emission spectroscopy (XES) have been widely used.[3–9] The transitions in Kα (2p → 1s) and Kβ mainline (Kβ$_{1,3}$) (3p → 1s) do not directly involve the valence orbitals, but are still sensitive to changes in valence electronic structure through the interactions between the valence electrons and the core/semi-core holes.[10–17]

Femtosecond XES combined with X-ray solution scattering (XSS) of the iron complex [Fe$^{II}$(bmip)$_2$]$^{2+}$ [bmip = 2,6-bis(3-methyl-imidazole-1-ylidine)-pyridine] showed similar femtosecond oscillations in the XES and XSS difference signals after valence excitation.[18] [Fe$^{II}$(bmip)$_2$]$^{2+}$ is a photosensitizer for solar energy conversion that can donate electrons upon photon absorption.[19,20] Photosensitizer efficiency is strongly linked to the excited state dynamics, which makes them primary targets for ultrafast spectroscopy, ranging from infrared to X-rays.[9,18,21–23] In [Fe$^{II}$(bmip)$_2$]$^{2+}$, the deduced photo-induced non-adiabatic dynamics is the following: (i) the excited molecular wavepacket decays non-radiatively from the singlet metal-to-ligand charge-transfer ($^1$MLCT) state and branches onto the triplet $^3$MLCT and triplet metal-centered ($^3$MC) states on a 100 fs time scale. (ii) Both triplet states has >1 ps lifetime and oscillatory structural dynamics occurs adiabatically on the $^3$MC potential energy surface (PES) with a 280 fs period. In this state, the coherent vibrational wavepacket moves along an iron–ligand stretching mode, which corresponds to changes in geometry along the ground state (GS) to $^3$MC distortion coordinate.[18] The presence of an oscillatory XES signal from this process, which matches the XSS structural data, shows that core-to-core XES can be sensitive to structural dynamics. However, although the K$\alpha$ difference signal showed a clear oscillatory pattern, no definitive assignment could be made for the corresponding mainline K$\beta_{1,3}$ signal.

K$\beta_{1,3}$ is usually more sensitive to valence electronic structure than K$\alpha$ because the radial distribution of a 3p hole overlaps directly with the valence orbitals.[13,17] K$\beta_{1,3}$ can thus provide detailed insights into metal–ligand bonding.[12–14,17,24,25] To understand the role of K$\beta$ XES in studies of excited state dynamics, it is important to analyze the structural sensitivity in more detail. To what extent is the sensitivity of mainline K$\beta$ XES higher than that of K$\alpha$ XES, and under which circumstances can the higher expected sensitivity overcome a lower emission intensity? Here these questions will be addressed by theoretically analyzing the sensitivity of K$\beta$ spectroscopy for an isolated process: pure structural dynamics in different electronic states of [Fe$^{II}$(bmip)$_2$]$^{2+}$.

A wide variety of different theoretical methods can be used to theoretically model X-ray processes, including multi-reference wavefunction methods, density-functional theory (DFT), and Green's function approaches.[26–32] The structural sensitivity of K$\alpha$ XES in [Fe$^{II}$(bmip)$_2$]$^{2+}$ has previously been analyzed using a multiconfigurational wavefunction model, based on the restricted active space (RAS) approach.[18,33] Here the same method will be used to analyze also the K$\beta$ signal to enable a side-by-side analysis of the two experiments. RAS is an *ab initio* multi-reference approach that gives a qualitatively correct description of the multiple open-shells associated with core–hole spectra of both ionic and covalent transition metal complexes.[25,30,34–37] This is important for modeling of K$\beta$ mainline XES signals of open-shell systems where the final state has strong multiplet interactions between the 3d electrons and the 3p hole,[13,25,38] and RAS has already been used to calculate K$\beta$ mainline XES.[25] In addition, RAS provides a chemically intuitive molecular orbital picture that directly connects to metal–ligand binding geometry. Another method that accurately describes multiplet interactions is the charge-transfer multiplet (CTM) model.[31,39] This model has previously been used to simulate iron L-edge resonant inelastic X-ray scattering (RIXS)[40,41] of [Fe$^{II}$(bmip)$_2$]$^{2+}$.[23] However, to analyze structural dynamics would require a well-defined mapping between metal–ligand distance and the charge-transfer mixing parameters.

One goal of spectral simulations is to aid the design of new experiments by predicting the emission energy regions that are most sensitive to structural dynamics. For this purpose, a cost-efficient X-ray simulation method would be advantageous. Among such methods are Green's function approaches,[32,42–44] widely used for solid-state systems, and DFT approaches for coordination complexes.[15,29,45–50] For K$\beta$ XES, these methods have, with few exceptions,[51] been used for valence-to-core K$\beta$ XES where the coupling between the open shells in the final state are weaker than in the K$\beta$ mainline.[15,52–55] DFT-based methods designed to qualitatively describe spectroscopy of open-shell coupling include DFT configuration–interaction (CI),[56] ligand-field (LF)-DFT,[48] and DFT restricted open-shell configuration interaction with singles (ROCIS).[47] Of these, only the latter has been used to study K$\beta$ (valence-to-core) XES. However, it is challenging to apply that method to systems with orbitally degenerate ground states, like the $^3$MC state.

Here the RAS results are compared with those from the one-electron orbital-energy approach in DFT (1-DFT).[15] Although that method, like many other DFT methods, does not give a formally correct description of the open-shell $^3$MC states, a major advantage is the relatively straightforward analysis. It has also already been applied in a range of applications.[15,46,57,58] Combining RAS and 1-DFT modeling for both K$\alpha$ and K$\beta$ XES allows for a thorough analysis of the relative sensitivity of the two experimental methods to coherent vibrations in [Fe$^{II}$(bmip)$_2$]$^{2+}$. It also makes it possible to outline the method requirements for simulations of the coupling between structural dynamics and XES observables.

## 2 Computational details

Simulations have been performed for eight different geometries along the GS – $^3$MC distortion coordinate. The geometries for the GS (1.940 Å average metal–ligand distance) and $^3$MC (2.066 Å) energy minima were optimized in a previous theoretical study, using PBE0/6-311G(d,p) in a MeCN polarizable continuum model (PCM).[19] Hybrid DFT functionals reproduce metal–ligand distances with a standard deviation around 0.02 Å.[59] Potential errors in the $^3$MC metal–ligand distances would not affect the calculation of structural sensitivity because, as will be shown below, that depends on the magnitude of the oscillations rather than the exact Fe–ligand distances. An additional six geometries were created on both sides of the $^3$MC energy minimum within 1.969–2.091 Å. These geometries include the turning points of the average oscillation in the wavepacket quantum dynamics simulations and XSS measurements,[18] with average Fe–ligand distances of −0.010 Å and +0.025 Å relative to the $^3$MC minimum. Cartesian coordinates of all structures are provided in the (ES). The GS structure belongs to the $D_2$ point group, but as the iron is six-coordinated, $O_h$ point group labels will be used to describe the metal orbitals.

1-DFT calculations were made with ORCA 5.0.1 using the def2-TZVP basis set.[60] Transition energies are calculated from energy differences of the core orbitals in the targeted state, calculations that can be directly performed for the lowest electronic states of a given spin multiplicity. An unrestricted Kohn–Sham formalism was used alongside the generalized gradient approximation (GGA) BP86 exchange–correlation functional, in line with previously published K$\beta$ simulations.[15] To compare the sensitivity of the results with respect to the choice of functional, calculations were also performed with the meta-GGA TPSS functional as well as the hybrid DFT B3LYP functional, although the orbital energy difference is not a well-defined approximation in the presence of Hartree–Fock (HF) exchange.[15] The effects of acetonitrile solvation on the 1-DFT results were included using the CPCM model.[61] Relative to the gas phase, this gives a very small 0.02 eV blueshift with negligible changes in spectral shape.

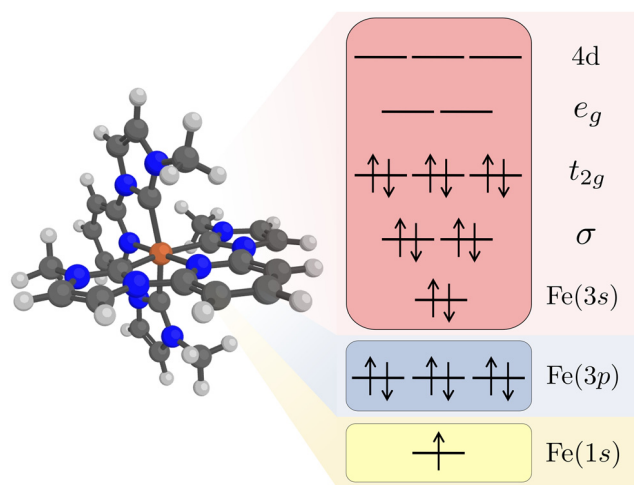

Fig. 1  Restricted active space (RAS) orbitals used in the simulations of Kβ X-ray emission. The electron configuration represents a 1s core–hole intermediate state with a closed-shell valence configuration. Note that the order of the Fe 3s and 3p orbitals are determined by the RAS spaces rather than the orbital energy order.

Kβ RAS calculations were made with OpenMolcas (v20.10).[62] Orbital optimizations were performed using the state-average RAS self-consistent field (SCF) formalism.[63,64] The RAS2 space, where all possible excitations are allowed, consists of 12 electrons distributed in 11 valence orbitals. These are the Fe 3s orbital, two Fe–ligand σ-bonding orbitals, three metal-dominated $t_{2g}$ orbitals, two metal-dominated $e_g$ orbitals of σ* character, and three metal 4d orbitals that can correlate with the $t_{2g}$ orbitals, see Fig. 1. Isodensity plots of the RAS2 orbitals are shown in Fig. S1 (ESI†). Compared to previously published RAS Kα XES simulations,[33] the Fe 3s orbital was added to the active space. This was required to get a stable active space for states with a 3p hole.

The Fe 1s orbital was placed in the RAS3 space, allowing for a maximum of two electrons, while the Fe 3p orbitals were placed in the RAS1 space, allowing for a maximum of one hole, see Fig. 1. K-edge X-ray emission simulations with RAS could previously only be performed for centrosymmetric complexes where excitations from s and p orbitals resulted in gerade and ungerade states respectively.[65,66] Here we use a projection operator, similar to the core-valence separation (CVS) technique,[67] to remove unwanted configurations with doubly occupied 1s and/or 3p orbitals.[68] Still, to ensure that the hole stays in the targeted orbitals, 1s and 3p orbitals were kept frozen in RASSCF optimizations.

Valence electronic states were calculated with singlet and triplet spin multiplicities. The $^3$MC state has a formal $t_{2g}^5 e_g^1$ orbital occupation, which gives rise to six close-lying states.[33] Unless otherwise mentioned, emission spectra are presented for the lowest of these states, with spectra for other states in the ESI.† 1s photoionisation from a triplet state leads to doublet (D) and quartet (Q) intermediate states. For the $^3$MC simulations, 9 doublet and 6 quartet 1s core–hole states were used, together with 50 doublet and 18 quartet 3p core–hole states. These are the same numbers as in the previous Kα XES simulations. RAS optimizations were performed separately for each spin multiplicity. To converge the large number of states, a dynamic configuration interaction (CI) algorithm was used that caps the total number of CI vectors but in each iteration allocates more vectors to states that have not yet been converged.[68] The converged RASSCF energies differ up to 0.1 eV depending on the choice of starting orbitals, indicating the presence of multiple local SCF minima. The final results are therefore obtained using the same set of starting orbitals for all geometries. The local minima have small effects on fitted values of total sensitivity, but adds some uncertainty in the quadratic fits of PESs as discussed below.

RAS calculations were performed using the double-zeta ANO-RCC-VDZP basis set,[69,70] and atomic compact Cholesky decomposition.[71] The effects of method choices on RAS X-ray simulations have been tested previously, and the effect of basis set size on spectral shape is limited.[72–74] RAS modeling was performed in gas phase. PCM solvent effects on RAS X-ray spectra are small as long as the ground electronic state remains the same,[74,75] in line with the above-mentioned 1-DFT results. The PCM model is sensitive to changes in the overall dipole moment, and metal-centered core transitions do not significantly change the dipole moment as the hole–electron pair is almost completely screened. Including dynamic correlation through second-order perturbation theory (RASPT2)[76] adds significant computational complexity but had limited effects for Kα sensitivity.[33] Dynamical correlation was important in the RAS modeling of resonant valence-to-core Kβ XES due to mixing between different valence states,[77] but as the semi-core 3p holes are low-lying the effect is smaller for the β mainline. Dynamical correlation has therefore not been included in neither Kα nor Kβ simulations.

Spin–orbit coupled (SOC) states were obtained using a Douglas–Kroll–Hess Hamiltonian and atomic mean field integrals[78,79] by the RAS state-interaction (RASSI) approach.[80,81] The RASSI method was also used to calculate electric dipole oscillator strengths of the emission processes. Intensities for the initial 1s photoionization were taken from the previous Kα modeling.[33] Orbital composition, as well as radial charge and spin densities, were calculated using the Multiwfn program.[82]

Calculated spectra were broadened using a 0.39 eV half-width-at-half-maximum (HWHM) Gaussian and a 0.81 eV Lorentzian function.[83] These values are the same as in the previous Kα simulations to facilitate the comparison between methods.[33] To align the calculated spectra, the GS spectra for the GS-minimum geometry were shifted to the Kβ emission maximum at 7057.9 eV. This gave corrections of −19.71 eV and 180.53 eV for RAS and 1-DFT BP86 respectively, similar in magnitude to other simulations with 1s core holes.[15,65,84] Intensities of the GS spectra were normalized to a maximum of 1.00, which gave scaling factors of 1/0.00570 for RAS and 1/2452 for 1-DFT BP86. To compare with the experimental data in ref. 18, the integrated sensitivity was also calculated for the full energy range 7056–7058.5 eV. Those calculations were normalized to a value of 1.00 for the GS spectra at the GS-minimum geometry, giving the normalization constants 1/0.0111 for RAS and 1/4967 for 1-DFT BP86. For 1-DFT BP86 Kα XES simulations, the GS spectra were aligned to the Kα emission maximum at around 6404.3 eV, which gave an energy shift of 160.7 eV and a scaling factor of 1/21954. The same energy shifts and intensity scaling factors were then used for all RAS, 1-DFT BP86 Kβ, and 1-DFT BP86 Kα spectra.

To connect the analysis to the commonly used atomic multiplet model, HF calculations of atomic ions have been performed using the CTM4XAS interface to an atomic multiplet code.[39,85] Note that these HF calculations have only been used to illustrate the atomic effects of different core holes, and not for any spectral simulations. As HF does not include electron correlation, electron–electron repulsion is overestimated and in the current calculations, dynamical electron correlation has been approximated by scaling down the exchange and Coulomb integrals between electron–electron pairs to 80% of their original value.

## 3 Results and discussion

In this section, Kβ simulations for the GS and $^3$MC equilibrium geometries are presented first, followed by the effects of structural dynamics. The differences in structural sensitivity of Kα and Kβ XES are then analyzed in detail, together with the origin of the structural sensitivity in RAS and 1-DFT. Finally, the possibility to predict experimental outcomes are discussed.

### 3.1 Kβ X-ray emission spectra

The experimental Kβ emission from the singlet ground state of $[Fe^{II}(bmip)_2]^{2+}$ has one main broad feature with an intensity maximum at 7057.9 eV, see Fig. 2 (black dotted line).[18] This maximum comes from the Kβ$_{1,3}$ mainline, associated with spin and electric–dipole allowed 3p to 1s transitions.

1s photoionisation from the closed-shell GS does not lead to any significant shake-up transitions in the RAS calculations, and only the lowest doublet intermediate state is considered. From there, the intense transitions lead to states with a 3p hole coupled to a closed-shell valence configuration. The Kβ$_{1,3}$ final states are split by deviations from formal $O_h$ symmetry and 3p spin–orbit coupling. After broadening, all final states appears within the same spectral envelope, see Fig. 2 (grey solid line). The 1-DFT BP86 spectrum also shows a main resonance, see Fig. 2 (grey dashed line), with a slightly more pronounced low-energy shoulder arising from the split of the 3p orbital energies. Compared to experiment the spectra are slightly too narrow. This is because the same spectral broadening has been used for all simulated spectra to facilitate a cross-method comparison, rather than adapting the broadenings to the specific conditions.

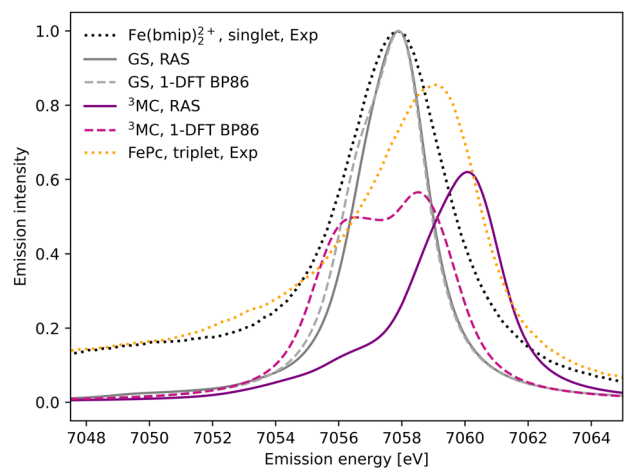

Fig. 2 Kβ X-ray emission spectra from experiment,[18] RAS and 1-DFT BP86 simulations. The simulated spectra for the electronic GS are calculated at the GS equilibrium geometry and the $^3$MC spectra at the $^3$MC equilibrium geometry.

As the $^3$MC state lacks an experimental steady-state spectrum, iron phtalocyanine (FePc) is used as the triplet reference, see Fig. 2 (yellow dotted line). This is a relevant model as population kinetics fitting of the $^3$MC XES difference spectrum shows strong similarities with the corresponding FePc data.[18] Compared to $[Fe^{II}(bmip)_2]^{2+}$, the main Kβ$_{1,3}$ feature is broader and shifted to higher energies. There is also a low-energy Kβ′ feature, split from the Kβ$_{1,3}$ by differences in exchange coupling between the open valence shell and the core hole. The RAS Kβ spectrum for the $[Fe^{II}(bmip)_2]^{2+}$ $^3$MC state, purple solid line in Fig. 2, shows similar behaviour as the triplet reference. This includes a significant blue-shift relative to the GS spectrum (2.2 eV), considerably more structure with a low-energy shoulder at 7056 eV, and lower maximum intensity. Separating the spectral contributions from quartet and doublet core-ionized states shows that the latter is dominant at lower energies, see Fig. S2 (ESI†). The emission spectrum from the second-lowest $^3$MC valence state is similar to that of the first, see Fig. S3 (ESI†).

The 1-DFT BP86 $^3$MC spectrum is different from the RAS one as it has two separate peaks of similar height separated by 2.0 eV, see Fig. 2 (purple dashed line). Compared to the GS, the high-energy peak is blue-shifted by 0.6 eV. The low-energy peak consists of transitions between states with holes in 1s and 3p α orbitals, corresponding to the RAS doublets, while the high-energy peak includes the corresponding transitions between β orbitals, corresponding to the RAS quartets. The large splitting is due to the strong 3p–3d exchange interactions. However, this two-peak structure is different from both the triplet RAS spectrum and the experimental reference spectrum, and most likely appears due to an incomplete treatment of the final-state multiplet effects.[15]

### 3.2 Structural sensitivity of Kβ XES

After photoexcitation the system evolves on the $^3$MC excited state surface and oscillates along a Fe–ligand stretching mode.

To model the sensitivity to structural dynamics, emission spectra are calculated at several geometries ranging from the GS minimum to the far side of the $^3$MC minimum, see Fig. 3. Starting with the $^3$MC RAS spectrum, there is a clear blue-shift with increasing the Fe–ligand distance. This blue shift appears in both doublet and quartet components of the $^3$MC simulations, see Fig. S4 (ESI†). The shape of the spectrum also changes slightly, with the low-energy side of the peak decreasing in intensity. RAS predicts a significant blue-shift with longer distances also for the closed-shell GS spectrum, see Fig. 3. This is accompanied by a decrease in the maximum intensity.

Shifts to higher energy are also seen in the 1-DFT BP86 spectra, see Fig. 3, although less pronounced than in the RAS simulations. Note that the two geometries with the shortest Fe–ligand distances are not included because here the calculations converge to triplet metal-to-ligand charge-transfer states ($^3$MLCT) state rather than $^3$MC states. 1-DFT BP86 predicts small changes in spectral shape for the $^3$MC state while the GS shows a small increase in maximum intensity with increasing distance. 1-DFT calculations with TPSS and B3LYP functionals give very similar spectra and spectral changes with distance, to those calculated with BP86, for both the $^3$MC state and the GS, see Fig. S5 and S6 (ESI†).

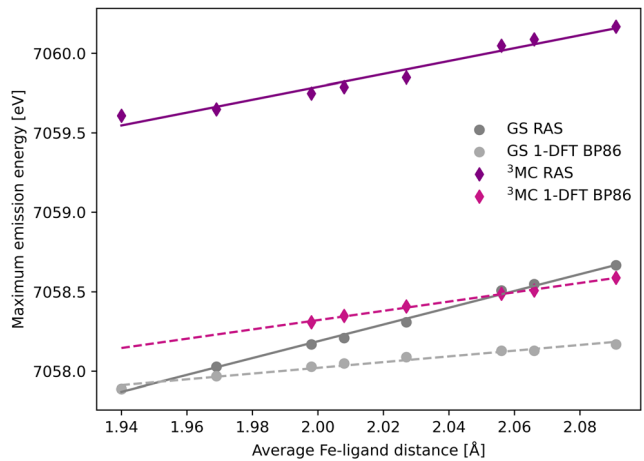

Fig. 4 Energies of the Kβ emission maxima for spectra in Fig. 3 plotted as a function of the Fe−ligand distance. Linear regression lines were fitted for each set of calculations.

The common denominator of all calculated spectra is a blue-shift with increasing the Fe–ligand distance. To quantify this effect, the energies of the RAS and 1-DFT BP86 Kβ emission maxima are plotted as a function of the Fe–ligand distance, see Fig. 4. This gives approximately linear energy-distance relations for all four sets of calculations. RAS gives a slope of 4.0 eV Å$^{-1}$ for the $^3$MC state and 5.3 eV Å$^{-1}$ for the GS. For the $^3$MC, both doublet and quartet intermediate states show similar blue-shifts, see Fig. S7 (ESI†), and so does the higher-lying $^3$MC states, see Fig. S8 (ESI†). As seen already from the spectra, the 1-DFT BP86 slopes are smaller, 2.9 and 1.8 eV Å$^{-1}$ for $^3$MC and GS respectively. The results with the other functionals are rather similar to those with BP86. For the $^3$MC state, TPSS and B3LYP gives slightly larger slopes than BP86, 3.3 and 3.7 eV Å$^{-1}$ respectively. For the GS they instead give slightly smaller slopes, 1.6 and 1.1 eV Å$^{-1}$, see Fig. S9 (ESI†). As the 1-DFT spectra and energy shifts show qualitative similarities, only the BP86 analysis will be described in full detail.

The femtosecond time-resolved Kβ experiment of [Fe$^{II}$(b-mip)$_2$]$^{2+}$ measured the difference in integrated intensity in the 7056–7058.5 eV energy range, see yellow area in Fig. 3. However, to facilitate a direct comparison to Kα XES, where the intensity was monitored at GS maximum, the intensity analysis is first performed for a single energy, the Kβ emission maximum at 7057.9 eV. With RAS the intensity decreases as the Fe–ligand distance increases, with gradients of −1.2 Å$^{-1}$ for the $^3$MC state and −2.2 Å$^{-1}$ for the GS, see Fig. 5. For $^3$MC the quartet contributions are more sensitive than the doublets, see Fig. S10 (ESI†), but the results from different $^3$MC states are similar, see Fig. S11 (ESI†). Instead looking at the 1-DFT BP86 calculations, they predict lower sensitivity for both the $^3$MC state (−0.4 Å$^{-1}$) and the GS (0.1 Å$^{-1}$).

To better understand the differences between the RAS and 1-DFT results, the sensitivity is divided into two different contributions: (i) energy shift sensitivity and (ii) spectral shape change. First, as the maxima of the $^3$MC emission spectra are blue-shifted with respect to the GS spectra, further blue shift with increasing metal–ligand distance ($r$) leads to a decrease in intensity ($I$), see Fig. 3. Assuming a constant spectral shape, the energy shift sensitivity ($\delta I(E)/\delta r$) can be expressed as:[33]

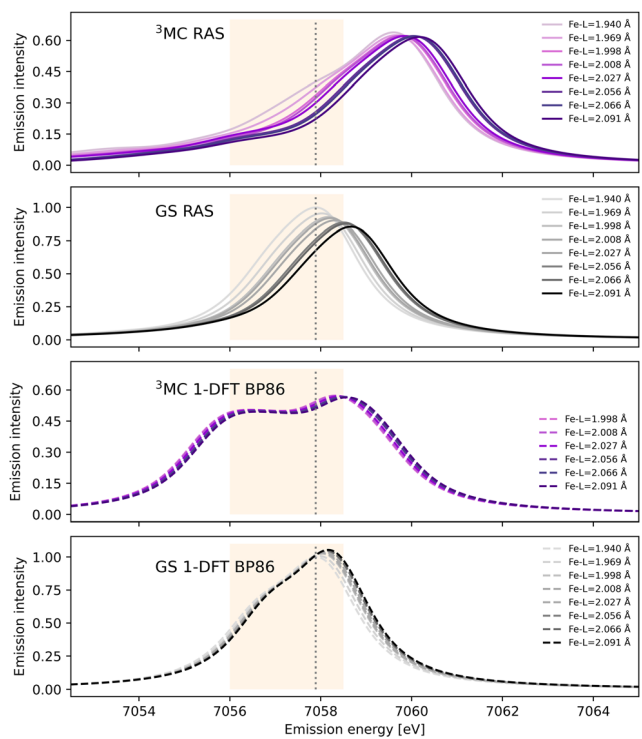

Fig. 3 RAS and 1-DFT BP86 simulated Kβ emission spectra for $^3$MC and GS electronic states. Spectra are calculated for up to eight geometries along the GS-$^3$MC distortion coordinate. At short Fe–ligand distances the $^3$MC state is not stable with respect to a triplet metal-to-ligand charge-transfer ($^3$MLCT) state and the corresponding spectra are not included. The grey dotted vertical line at 7057.9 eV represents the maximum emission intensity of the experimental GS spectrum. The yellow area represents the 7056–7058.5 eV energy range used for the Kβ difference spectrum in ref. 18.

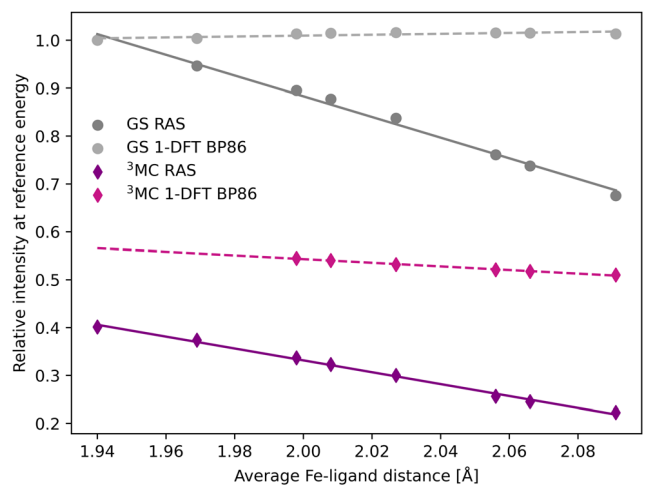

Fig. 5 Emission intensities at the GS Kβ maximum energy 7057.9 eV for XES spectra in Fig. 3 plotted as a function of the Fe–ligand distance. Linear regression lines were fitted for each set of calculations.

$$\frac{\delta I(E)}{\delta r} = \frac{\delta I(E)}{\delta E_{max}} \frac{\delta E_{max}}{\delta r} \quad (1)$$

where $\delta E_{max}/\delta r$ (energy shift) is the gradient of the maximum emission energy with distance, see Fig. 4, and $\delta I(E)/\delta E_{max}$ (intensity gradient) is the change in intensity caused by the shift. For small shifts, this equals the negative slope of the intensity at a given energy. The numerical values of these factors, evaluated at the 7057.9 eV reference energy and the $^3$MC minimum geometry, are shown in Table 1.

In most cases, there are significant differences between the total sensitivity, extracted from Fig. 5, and the energy shift sensitivity, see Table 1. This can largely be attributed to changes in spectral shape with Fe–ligand distance, see Fig. 3. As an example, for the $^3$MC RAS calculation the energy shift sensitivity is $-0.64$ Å$^{-1}$ compared to a total sensitivity of $-1.24$ Å$^{-1}$. The difference is due to a suppression of the $^3$MC low-energy shoulder with increasing distance, a feature that is located in the same energy region as the GS maximum.

Although dynamics is not observed for the GS in the experiment, comparing the RAS results for two different states provides more insight into potential common origins of the observed sensitivity. The GS shows a higher total sensitivity ($-2.16$ Å$^{-1}$) compared to $^3$MC ($-1.24$ Å$^{-1}$), see Table 1. The most important factor is the intensity gradient, twice as large for the GS. This is due to a less structured (narrow) Kβ$_{1,3}$ peak with a large slope at the reference emission energy for the longer Fe–ligand distances, see Fig. 3. Together with a slightly larger energy shift, this gives a high energy shift sensitivity ($-1.75$ Å$^{-1}$). There is also a decrease in maximum intensity of the GS with Fe–ligand distance, which further lowers the intensity at the reference energy and contributes to the total sensitivity.

The 1-DFT BP86 calculations predict lower sensitivities compared to RAS. For the $^3$MC state, the total sensitivity is $-0.38$ Å$^{-1}$, three times lower than the RAS value. The difference is even larger for the GS, where 1-DFT even predicts a small increase in intensity, see Table 1. Both energy shifts and intensity gradients are smaller with 1-DFT BP86 compared to RAS, which gives significantly reduced energy shift sensitivities. The shape of the $^3$MC spectrum does not change much, but the GS maximum intensity increases with distance and this cancels the effect of the energy shift.

The Kβ sensitivity results do not change significantly if the analysis is instead performed for the entire energy region used in the time-resolved experiment, see Fig. S12 (ESI†). This energy range covers the low-energy side of the RAS-calculated $^3$MC spectrum, see Fig. 3, and a blue-shift with increasing Fe–ligand distance leads to a decrease in intensity at almost all points within the specified energy range. With RAS the total integrated sensitivity ($\delta I_{area}/\delta r$) are $-1.1$ Å$^{-1}$ and $-2.6$ Å$^{-1}$ for the $^3$MC and GS respectively, with much lower sensitivity with 1-DFT BP86 ($-0.4$ Å$^{-1}$ and 0.0 Å$^{-1}$), see Table 1.

### 3.3 Comparing sensitivity of Kβ and Kα XES

In the [Fe$^{II}$(bmip)$_2$]$^{2+}$ experiment, the XES oscillations were more clearly seen in the Kα signal compared to the Kβ one.[18] To better understand the relative sensitivity of the two signals, the current Kβ results are compared to previously published RAS Kα results,[33] and new Kα results from 1-DFT BP86 modeling. Note that RAS Kα and Kβ calculations are done with slightly different active spaces, with an additional 3s orbital in the Kβ calculations, see computational details.

Table 1 Sensitivity of Kα and Kβ spectra to changes in the Fe–ligand distance. The intensity gradient ($\delta I/\delta E_{max}$) is calculated at the energy of the GS emission maximum (7057.9 eV) for the $^3$MC minimum geometry (Fe–L = 2.066 Å). The Energy shift is extracted from Fig. 4. Energy shift sensitivity ($\delta I/\delta r$) is calculated from eqn (1) and total sensitivity is extracted from Fig. 5. Both values refer to the intensity at the GS emission maximum. Total integrated sensitivity ($I_{area}/\delta r$) refers to the normalized integrated intensity between 7056–7058.5 eV, which is only measured for Kβ XES, the data is extracted from Fig. S12 (ESI). All RAS Kα values are taken from ref. 33

| | | RAS | | | | 1-DFT BP86 | | | |
| | | $^3$MC | | GS | | $^3$MC | | GS | |
| Factor | Expression | Kα[33] | Kβ | Kα[33] | Kβ | Kα | Kβ | Kα | Kβ |
| --- | --- | --- | --- | --- | --- | --- | --- | --- | --- |
| Intensity gradient | $\delta I/\delta E_{max}$ (eV$^{-1}$) | −0.21 | −0.16 | −0.52 | −0.33 | −0.24 | −0.10 | −0.33 | −0.24 |
| Energy shift | $\delta E_{max}/\delta r$ (eV Å$^{-1}$) | 3.0 | 4.0 | 3.3 | 5.3 | 1.4 | 2.9 | 1.5 | 1.8 |
| Energy shift sensitivity | $\delta I/\delta r$ (Å$^{-1}$) (product) | −0.63 | −0.64 | −1.72 | −1.75 | −0.34 | −0.29 | −0.50 | −0.43 |
| Total sensitivity | $\delta I/\delta r$ (Å$^{-1}$) (fit) | −0.61 | −1.24 | −1.51 | −2.16 | −0.38 | −0.38 | 0.08 | 0.09 |
| Total integrated sensitivity | $\delta I_{area}/\delta r$ (eV Å$^{-1}$) (fit) | — | −1.17 | — | −2.63 | — | −0.24 | — | −0.01 |

The equilibrium geometry Kα spectra are shown in Fig. S13 (ESI†). As for Kβ, the $^3$MC Kα spectra are blue-shifted relative to the GS, and there is a significant drop in maximum intensity. Spectra for different Fe–ligand distances are shown in Fig. S14 (ESI†). The results are in many aspects qualitatively similar to Kβ, with near linear blue-shifts of the emission maxima with increasing distance, see Fig. S15 (ESI†). However, in all simulations energy shifts are smaller for Kα compared to Kβ, see Table 1 and Fig. 6. As an example, with RAS the $^3$MC energy shift is 3.0 eV Å$^{-1}$ for Kα XES compared to 4.0 eV Å$^{-1}$ for Kβ.

The smaller energy shifts in Kα are offset by larger intensity gradients caused by narrower spectral shapes. The energy shift sensitivity is therefore similar for Kα and Kβ, see Table 1. The narrower Kα line shapes are due to smaller exchange couplings between valence electrons and the 2p hole in the Kα final state, while the 2p spin–orbit coupling is so large that it forms completely separate Kα$_1$ and Kα$_2$ peaks, see Fig. S13 (ESI†).

Despite the similar energy shifts, RAS still predicts a higher total sensitivity for Kβ, see Fig. 7. Taking the $^3$MC results as an example, the sensitivity is twice as high for Kβ (−1.24 Å$^{-1}$) compared to Kα (−0.61 Å$^{-1}$), see Table 1 and Fig. S16 (ESI†). This is due to large changes in the $^3$MC Kβ spectral shape with geometry, as discussed above, which are not observed in the Kα simulations, see Fig. S13 (ESI†). The GS results shows a similar increase in total sensitivity when going from Kα to Kβ, but the relative increase is smaller because the sensitivity is already high to begin with. This can be contrasted with the results from 1-DFT BP86, where Kα and Kβ have similar, but also rather low, sensitivity.

To summarize, there are three factors that influence the different sensitivity of Kβ mainline and Kα XES: the size of the energy shift, the slope of the emission curve at a given energy, and the change in spectral shape. Compared to Kα XES, Kβ has a smaller slope, but this is overcome by a larger energy shift and larger changes in spectral shape. The connection between these factors and electronic structure will be analyzed in more detail below.

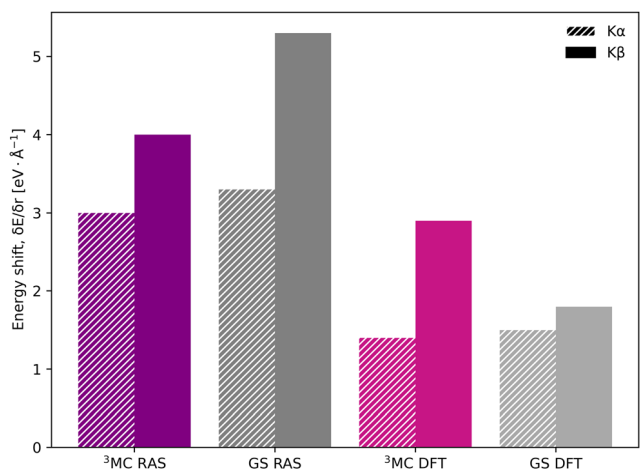

Fig. 6 Energy shifts ($δE_{max}/δr$) in eV Å$^{-1}$ for Kα (striped bars) and Kβ (solid bars) spectra, calculated with RAS and 1-DFT BP86 for $^3$MC and GS electron configurations.

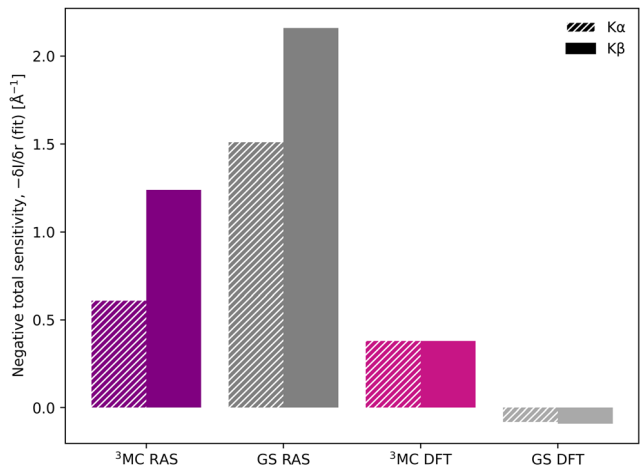

Fig. 7 Sensitivity of Kβ and Kα to changes in the Fe–ligand distance, shown as the negative value of the total sensitivity ($−δI/δr$) at the GS emission maximum energy (6404.3 eV for Kα and 7057.9 eV for Kβ).

### 3.4 Rationalizing structural sensitivity in RAS

To rationalize the structural sensitivity, it is valuable to look into the interactions between the valence and the different core/semi-core orbitals. This is first done for RAS, where the core–hole states are explicitly calculated, followed by 1-DFT where only the initial states are available.

RAS predicts spectral blue-shifts with increasing Fe–ligand distance for both Kα and Kβ XES. In Kα XES, this effect was assigned to a shift of the minimum energy metal–ligand bond distance between 1s and 2p core–hole states due to differences in valence orbital contraction.[18,33] A similar analysis of the changes in the radial charge densities (RCD) for Kβ XES is shown in Fig. 8.[86] It shows a significant contraction of the 3d-type orbitals in the 1s-hole state compared to the $^3$MC valence state, and a much smaller contraction in the final 3p-hole state due to the spatial overlap of the 3p and 3d orbitals. The effects of 1s and 3p holes are consistent with the 3d orbital energies in the atomic iron(II) model, see Table S1 (ESI†).[85]

In a σ-donor complex, stabilization of the 3d metal levels decreases the energy difference to the low-lying filled ligand orbitals and σ-orbital covalency is therefore higher in the 1s-hole state, see Table 2. As for the RCD, the covalency of the 3p-hole state is relatively close to the $^3$MC state. However, the relative ordering of the two changes with geometry, with the 3p-hole state being more covalent at the $^3$MC minimum geometry and slightly less covalent at the GS geometry.

The core–hole effects on metal–ligand bonding can be connected to the blue-shift of the emission energies through the respective PESs of the different states. After the optical excitation, the complex moves coherently on the $^3$MC PES with a 280 fs period. Due to the short lifetime (1 fs) of the 1s core hole, the Kβ/emission process can be approximated as vertical transitions from a 1s-hole PES to a 3p-hole PES, see Fig. 9. As explained in detail in ref. 33, these PES have different shapes due to changes in metal–ligand bonding, which makes emission energies geometry dependent. Compared to the ground state, the creation of a 1s core leads to a contraction of the metal 3d levels, see Fig. 8, and a corresponding lowering of the energy levels, see Table S1 (ESI†). In a two-configuration model, where these metal levels interact with filled ligand levels, this leads to increased metal–ligand covalency, see Table 2. Geometrically, the result is a shortening of the metal–ligand bonds in the 1s core–hole state. As the perturbation is smaller for the 3p hole, see Fig. 8, the metal–ligand distances in this state are expected to be more similar to the ones in the ground state.

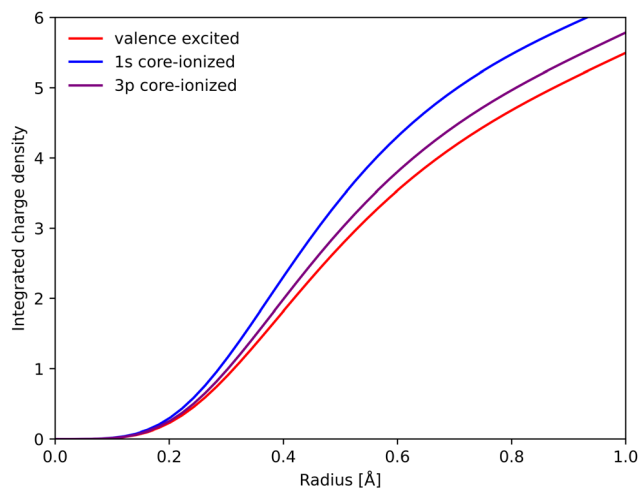

Fig. 8 Integrated RAS radial charge density (RCD) plots of the $^3$MC valence state, and the corresponding states with 1s and 3p holes calculated at the $^3$MC equilibrium geometry. To avoid effects of changes in core orbital occupation, only the Fe 3d $t_{2g}$ and $e_g$ orbitals were included. Radial spin densities (RSD) for the same states are shown in Fig. S17 (ESI†).

Table 2 RAS orbital covalency analysis for selected molecular orbitals, see Fig. 1, in three different states at the GS and $^3$MC minimum geometries. Values represent the percent of metal contributions to the molecular orbitals. For metal-dominated $t_{2g}$ and $e_g$, lower values correspond to stronger mixing between metal and ligand orbitals and thus stronger covalency. For the ligand-dominated σ orbitals stronger metal–ligand mixing instead leads to larger numbers. The covalencies of the $^3$MC and $^3$MC 1s states differ slightly from those in ref. 33 due to differences in the RAS2 space

| State | GS geometry | | | $^3$MC geometry | | |
|---|---|---|---|---|---|---|
| | σ | $t_{2g}$ | $e_g$ | σ | $t_{2g}$ | $e_g$ |
| $^3$MC | 18.7 | 92.7 | 76.3 | 9.4 | 91.8 | 89.7 |
| $^3$MC 1s hole | 22.6 | 92.8 | 73.2 | 15.6 | 95.8 | 79.4 |
| $^3$MC 3p hole | 16.3 | 94.9 | 81.4 | 10.8 | 96.7 | 87.1 |

Using the PES analysis, it is possible to qualitatively rationalize the larger energy shift in Kβ (4.0 eV Å$^{-1}$) compared to Kα (3.0 eV Å$^{-1}$). Assuming that the PESs are quadratic with similar force constants, the shift in emission energy is proportional to the force constant ($k$) and the distance between the minima ($\Delta r$).[33] The $^3$MC valence state minimum is at 2.11 Å, which decreases to 2.06 Å in the 1s core–hole state and goes back to 2.11 Å for the 3p-hole PES. This gives a difference ($\Delta r$) of approximately 0.05 Å between the relevant states in Kβ emission, while in Kα 1s and 2p states were separated by approximately 0.02 Å.[33] Although the PES fitting values are relatively uncertain due to the presence of close-lying local minima in the RASSCF optimization, the larger displacement in Kβ should lead to a larger energy shift. That effect is slightly moderated by a smaller force constant for the 3p-hole state compared to the 1s-hole state, at least for Fe–ligand distances below the 3p-hole minimum. The difference in force constants should introduce a non-linear component to the energy shift, but the small distance range makes it difficult to isolate such a contribution.

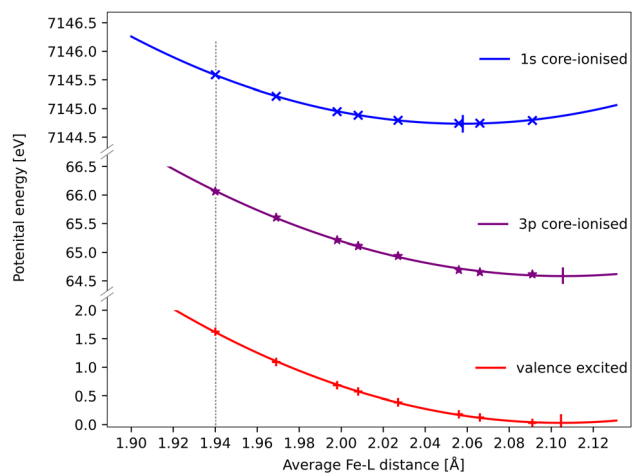

Fig. 9 RAS energies as a function of the Fe–ligand distance for the $^3$MC valence state, the lowest $^3$MC 1s-hole state, and the $^3$MC 3p-hole state with the highest transition probability from that 1s-hole state. The grey dotted vertical line at 1.940 Å represent the average Fe–L bond length for the GS geometry. Quadratic PES fits give minima at 2.06 Å, 2.11 Å and 2.11 Å for the 1s, 3p core–hole and $^3$MC valence states and corresponding force constants of 122 eV Å$^{-2}$, 108 eV Å$^{-2}$ and 117 eV Å$^{-2}$. the fit values must be considered uncertain considering deviations from quadratic behaviour, lack of data points close to the minima, and the presence of multiple local SCF minima. The force constant for the 1s hole is smaller than reported in ref. 33, likely due to a stabilization of longer Fe–ligand distances by the additional 3s orbital.

In this analysis, the sensitivity of the Kβ energies is not explained as an effect of a strong perturbation caused by the 3p hole. In fact, creation of the 3p hole results in relatively similar descriptions of metal–ligand interactions as in the valence state. The large effects instead comes from the larger perturbation caused by the deep-lying core hole. RAS should predict positive shifts for complexes where the 1s PES minimum occurs at shorter distance compared to the 3p PES. As the 1s hole lowers the metal-centered orbitals relative the ligand orbitals, this happens for systems where higher-lying metal orbitals interacts with lower-lying ligand orbitals. This is the case for systems dominated by ligand (σ) donation.

When it comes to the shape of the Kβ spectra, the splitting between Kβ$_{1,3}$ and Kβ′ lines can be correlated to metal–ligand covalency.[12–14,17,24,25] Lower covalency leads to more localized 3d orbitals, larger 3d–3p exchange interactions and thus larger splittings. For high-spin iron systems, a linear relation between ground state covalency values and the Kβ mainline splitting could be observed.[25] The same trend for this system would lead to a small increase in the splitting with metal–ligand distance. This can be seen in the 1-DFT BP86 spectra where the splitting between the two peaks increases (by 1.2 eV) along the full distance range. However, in RAS the Kβ′ features are weak for both closed-shell GS and $^3$MC states, which makes it difficult to quantitatively analyze such effects.

### 3.5 Rationalizing structural sensitivity in 1-DFT

All spectra calculated with 1-DFT blue shift with increasing Fe–ligand distance. However, the energy shift is generally smaller than with RAS, see Fig. 6. In 1-DFT core–hole states are not explicitly modelled, and the emission energy is approximated by the orbital energy difference, with Kβ = $\varepsilon_{3p} - \varepsilon_{1s}$ and Kα = $\varepsilon_{2p} - \varepsilon_{1s}$. The energy shift in 1-DFT can thus be analyzed by plotting the relevant orbital energies as a function of metal–ligand distance, see Fig. 10. The 1s orbital energy decreases with distance, 2p also decreases but with a slower rate, while the 3p orbital energy increases. This leads to increasing energy difference with distance for both Kα and Kβ XES, with the largest effect in Kβ.

Numerically, linear fits of the BP86 orbital energies give gradients of −1.76 eV Å$^{-1}$ for 1s and 0.34 eV Å$^{-1}$ for 3p, which corresponds to a Kβ energy shift of 2.10 eV Å$^{-1}$. The same analysis for the 2p orbitals gives a gradient of −0.39 eV Å$^{-1}$, which combined with the 1s orbital energy gives a total Kα energy shift of 1.37 eV Å$^{-1}$. Compared to the values from the XES spectra in Fig. 6, the Kβ shift is underestimated while the Kα shift matches well.

The energy changes of the 1s and 2p orbitals can be analyzed using an atomic model because they are not directly involved in bonding.[87] The stabilization of these core orbitals with increasing the Fe–ligand distance corresponds to a decrease in electron density at the iron center. In [Fe$^{II}$(bmip)$_2$]$^{2+}$, bonding is dominated by strong σ donation from the occupied ligand orbitals. Increasing the Fe–ligand leads to a loss of σ donation, see Table 2, and reduces the penetration of the σ-bonding electrons to the iron.[88] Pairwise electron–electron interactions are strongly affected by the small region around the nucleus. Valence electrons have larger repulsive interactions with 1s compared to 2p due to their close proximity to the nucleus and the absence of a radial node, see Table S2 (ESI†).[87,89,90] This leads to larger stabilization of 1s compared to 2p, and thus increasing 2p → 1s emission energies.

Electrostatic arguments have been less effective for Kβ. Here the 3p–3d exchange interactions are dominant, and the position of the Kβ$_{1,3}$ peak mainly correlates with the number of unpaired 3d spins.[13,17] The interactions between the metal 3p orbitals and the ligand also complicates the picture. Here the Fe 3p orbitals increase in energy for the shortest bond lengths and then approaches an asymptotic behaviour at longer geometries. With 1-DFT the shift to positive energies occurs for systems where the electron density decreases at the nuclei with increasing metal–ligand distance, which should be affected by the amount of net donation between metal and ligand. The connection between positive shift and net ligand donation thus appears in both RAS and 1-DFT modeling.

To rationalize the differences between the RAS and 1-DFT calculations, it is useful to consider two separate final-state effects: (i) the direct effect of the core hole on the metal–ligand bonding, and (ii) the multiplet effects. The first effect can be isolated from the calculations of the singlet GS, because these spectra are not affected by any multiplet effects. Here the 1-DFT energy shift, 1.1–1.8 eV Å$^{-1}$ is consistently smaller than the RAS value (5.3 eV Å$^{-1}$). This supports the idea that 1-DFT underestimates the energy shift because it does not fully account for the effect that the core hole has on the metal–ligand bonding. The second effect is relevant only for the final state of the $^3$MC spectrum, where the lack of multiplet effects can lead to inaccurate descriptions of both spectral shape and position. The analysis of the $^3$MC results is therefore more complicated. Here the high-energy 1-DFT peak shifts by 2.9–3.7 eV Å$^{-1}$, which is only slightly lower than the 4.0 eV Å$^{-1}$ obtained in RAS. This relatively large 1-DFT shift comes from an overall blue-shift, combined with an increased splitting of the two spectral peaks. Instead looking at the low-energy peak, these shifts are lower (1.7–2.5 eV Å$^{-1}$). In the $^3$MC state, the large peak splitting in the 1-DFT spectra thus partially compensates for a smaller overall blue-shift.

### 3.6 Optimizing Kβ sensitivity

The spectral simulations could ideally be used to design future Kβ XES studies that probes electronic and structural dynamics. This will be illustrated for a general model considering two isolated processes: (i) electronic dynamics, referring to transitions between electronic states 1 and 2, and (ii) coherent structural dynamics in state 2. Detection of electronic dynamics requires a difference in Kβ XES intensity between the two states. Using the $[Fe^{II}(bmip)_2]^{2+}$ GS (state 1) and $^3MC$ (state 2) as example, that intensity difference is plotted as a function of emission energy in Fig. 11. This difference is large around the emission maximum of each of the two electronic states, 7057.9 eV and 7060.1 eV in RAS modeling. The sensitivity to electronic dynamics disappears when the two states have the same emission intensity, which in the RAS calculation occurs at 7059.4 eV.

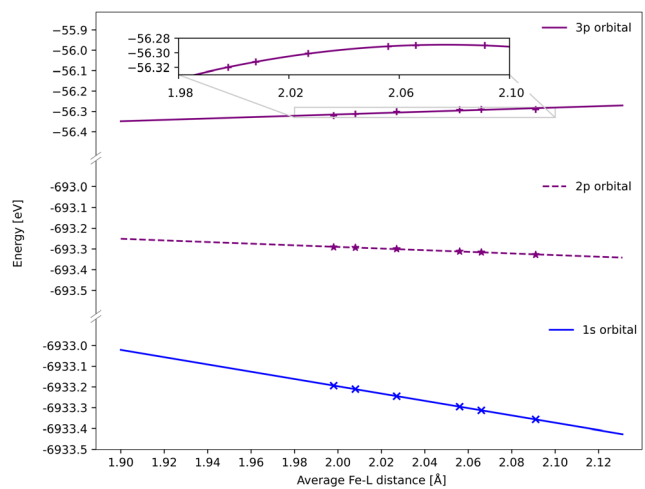

Fig. 10 DFT BP86 orbital energies of the 1s, 2p and 3p orbitals along the GS minimum – $^3$MC minimum coordinate for the $^3$MC state. For the p-orbitals the energy is an average of all three orbitals with the same principal quantum number. Linear regressions give gradients of −1.76 eV Å$^{-1}$ for 1s, −0.39 eV Å$^{-1}$ for 2p, and 0.34 eV Å$^{-1}$ for 3p, although the 3p orbital energies deviate from linear behavior (see inset).

Instead assuming that the only time-dependent process is structural dynamics in state 2, the intensity difference comes from the geometrical oscillations around the turning points in the coherent wavepacket dynamics, with average Fe–ligand distances of −0.010 Å and +0.025 Å relative to the $^3MC$ minimum. This difference is high for emission energies with a high gradient of the XES curve, see Fig. 11 (purple curve, right $y$-axis). For RAS, this occurs on both sides of the XES maximum, around 7058.2 eV and 7060.9 eV. In this particular case, high structural sensitivity in state 2 occurs close to the emission maximum of state 1, but the relative sensitivity to electronic and structural dynamics can still be modulated by changing emission energy. Different methods for estimating Kβ structural sensitivity, including the total sensitivity, gives similar results, see Fig. S18 (ESI†). The corresponding Kα results are shown in Fig. S19 (ESI†).

The analysis in Fig. 11 assumes a 100% excitation yield and a complete separation of electronic and structural dynamics. In real pump–probe experiments, incomplete yields and coupling between different degrees of freedom will lead to a more complicated picture. In the present simulations, the XES spectra will be sensitive to structural dynamics in both states. Coherent structural dynamics can thus be detected in any state, with the unlikely exception where two excited states vibrate completely out of phase.

Comparing the RAS and 1-DFT BP86 results in Fig. 11, the overall structure is the same because both predict a spectral blue-shift with increasing distance. However, when looking in more detail some important differences appear. This includes the overall lower structural sensitivity in the 1-DFT modeling as shown above for the GS maximum emission energy, see Fig. 7. The differences in spectral shape also leads to qualitative differences in some regions, most prominently between the two peaks in the 1-DFT $^3MC$ spectrum. There the local minimum gives a very small intensity gradient, and the 1-DFT total sensitivity becomes close to zero. In the RAS spectrum, that energy is on the left side of the peak with a relatively large intensity gradient and total sensitivity.

If the emission difference signal is obtained by integrating over an energy range, that range can also be chosen to optimize the sensitivity, see Fig. S20 (ESI†). If the intensity difference mainly depends on an energy shift, integrating over an energy range centered on the emission maximum would largely cancel positive and negative intensity contributions on each side of the peak. This is equivalent to measuring a single energy at the emission maximum where the gradient of the XES curve is small. In the $[Fe^{II}(bmip)_2]^{2+}$ XES experiment, the energy range was instead centered on the left of the $^3MC$ maximum, an area with large sensitivity as almost all regions of the spectrum gives a negative intensity contribution, see Fig. 3. The difficulty to detect structural dynamics in Kβ XES of $[Fe^{II}(bmip)_2]^{2+}$ was thus not due to a sub-optimal collection regime.

The RAS and 1-DFT predictions for sensitivity to dynamics cannot be directly compared to experiment, but their reliability can be judged by method considerations and static reference spectra. Starting with the energy shift, RAS predicts a larger value than 1-DFT. As the shift is strongly affected by core–hole effects on the metal–ligand bonding, which are explicitly included only in RAS, it is reasonable to assume that 1-DFT would underestimate this effect in many situations. When it comes to spectral shape, the GS emission process should be described equally well by the single and multi-configurational methods as it only includes one open shell. For the $^3MC$ states, RAS gives a qualitatively correct description of the multiple open shells and the spectral shape resembles the triplet reference spectrum. Considering the large effect of spectral shape on total sensitivity, using the spectral shapes from 1-DFT calculations can lead to incorrect conclusions. To get quick predictions, a possibility is then to use 1-DFT to estimate the direction of the energy shift, and then to use a single-geometry reference spectrum, *e.g.*, from a multi-configurational approach, to generate a more accurate spectral shape.

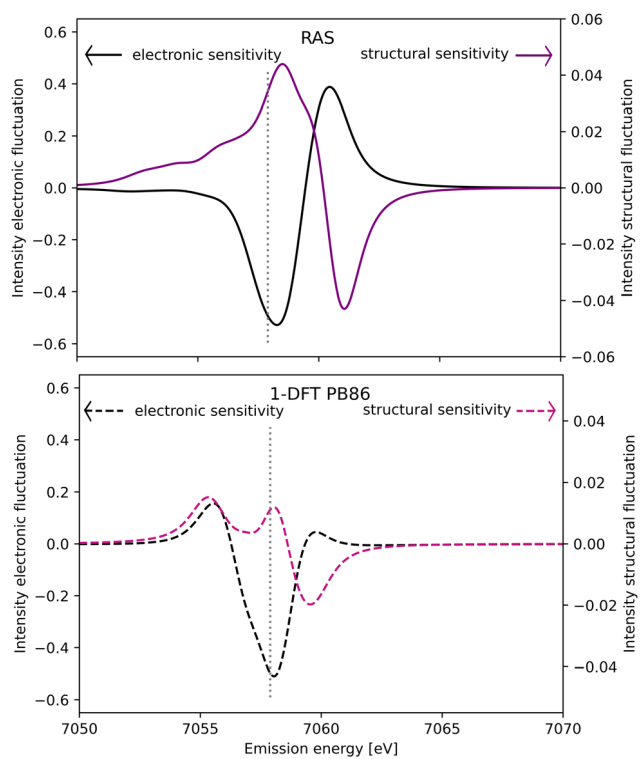

Fig. 11 Sensitivity of difference Kβ XES spectra to electronic (black line, left $y$-axis) and structural dynamics (purple line, right $y$-axis) as a function of the emission energy. Electronic dynamics represent transitions between two different electronic states (at a given geometry) and electronic sensitivity is calculated as the difference in intensity between $[Fe^{II}(bmip)_2]^{2+}$ GS and $^3MC$ electronic states at the $^3MC$ equilibrium geometry. The structural sensitivity represents the intensity difference between the two geometries that represent the turning points in wavepacket simulations (Fe–L distances of 2.056 and 2.091 Å).[18] Results are shown for RAS (solid lines, top figure) and 1-DFT BP86 (dashed lines, bottom figure). The grey dotted vertical line at 7057.9 eV represents the maximum emission intensity of the experimental GS spectrum.

## 4 Conclusions

The sensitivity of $K\beta_{1,3}$ XES to structural dynamics in the excited state of an iron photosensitizer is due to a blue-shift of the emission energy with increasing metal–ligand distance. This in turn leads to a change in intensity at a given energy or energy range. The result supports the suggestion that the $K\beta$ XES difference signal in $[Fe^{II}(bmip)_2]^{2+}$ shows a femtosecond oscillation due to coherent wavepacket dynamics. In the RAS calculations, the energy shift comes from differences in the PES minima between 1s and 3p core–hole states. The computationally more efficient 1-DFT orbital energy approximation also shows blue-shifts of the emission energies with increasing metal–ligand distance. However, the shifts are smaller than with RAS and the predicted sensitivities are also smaller.

$K\beta_{1,3}$ is more sensitive to structural dynamics compared to $K\alpha$ XES, with RAS predicting a factor of two for the $^3$MC state. The shift is affected by three factors: the size of the energy shift, the slope of the emission curve at a given energy, and changes in spectral shape with distance. $K\beta$ sensitivity is favored by a more sensitive spectral shape, and a larger energy shift. The latter effect is due to a larger displacement of the initial and final state PES minima in $K\beta$ compared to $K\alpha$. The drawback of using $K\beta_{1,3}$ is an approximately ten times lower photon count.

The simulations show a strong effect of the selected emission energy, or energy range, on the measured sensitivity. The calculations can therefore support the design of experiments studying the interplay between ultrafast electronic and structural dynamics in excited states.

## Author contributions

JR: investigation, validation, formal analysis (RAS calculations), visualization, writing – first draft. RS: investigation, formal analysis (DFT calculations). Writing – first draft. MV: methodology, formal analysis, writing – reviewing and editing. ML: conceptualization, formal analysis, writing – first draft. Writing – reviewing and editing.

## Conflicts of interest

There are no conflicts to declare.

## Acknowledgements

We thank Kristjan Kunnus and Kelly Gaffney for valuable comments on the manuscript draft, and thank Oscar Sommerbo for comments on the revised manuscript. ML acknowledges support from the Swedish Research Council (grant 2022-04794). The computations were enabled by resources provided by the Swedish National Infrastructure for Computing (SNIC) at the National Supercomputer Centre at Linköping University (Tetralith), partially funded by the Swedish Research Council through grant agreement no. 2018-05973. The project is funded by the European Union through ERC grant 101040356 (MV). Views and opinions expressed are however those of the author(s) only and do not necessarily reflect those of the European Union or the European Research Council Executive Agency. Neither the European Union nor the granting authority can be held responsible for them.

## Notes and references